\documentstyle[11pt,paspconf]{article}

\input epsf.sty

\def\kms {km~s$^{-1}$}
\def\lsim{ \lower .75ex \hbox{$\sim$} \llap{\raise .27ex \hbox{$<$}} }
\def\gsim{ \lower .75ex \hbox{$\sim$} \llap{\raise .27ex \hbox{$>$}} }

\begin{document}

\title{Dark Matter Halos and Disk Rotation Curves}

\author{Julio F. Navarro\altaffilmark{1}}
\affil{Steward Observatory, The University of Arizona, Tucson, AZ 85721.}


\altaffiltext{1}{Bart J. Bok Fellow}



\begin{abstract}
Cosmological N--body simulations have revealed a remarkable similarity
in the structure of dark matter halos formed in hierarchically
clustering universes. Regardless of halo mass, cosmological
parameters, and power spectrum of initial density fluctuations, the
spherically averaged density profiles of dark matter halos have a
universal shape. The logarithmic slope of this profile is shallower
than isothermal near the center, and steepens gently outwards,
becoming steeper than isothermal near the halo virial radius. This
profile can be well approximated by a simple formula with only two
free parameters: halo mass and ``characteristic'' density, e.g., the
density at the radius where the logarithmic slope equals the
isothermal value. This characteristic density is proportional to the
mean density of the universe at the time of collapse of each system,
and decreases systematically with increasing halo mass, reflecting the
later collapse of more massive halos. I use these results to examine
what constraints can be derived for Cold Dark Matter models from the
rotation curves of disk galaxies.
\end{abstract}


\keywords{globular clusters,peanut clusters,bosons,bozos}


\section{Introduction}

The structure of dark matter halos formed through gravitational
collapse in hierarchically clustering universes has received close
attention ever since the work of Gunn \& Gott (1972) showed that the
virialized structure of halos may contain clues to the cosmological
parameters. Subsequent analytic work, which focussed mainly on the
density profiles of systems formed from scalefree initial conditions,
concluded that the equilibrium mass profiles of dark halos should be
well approximated by power laws, and that the power-law slope should
depend sensitively on the cosmological parameters (Fillmore \&
Goldreich 1984, Bertschinger 1985). These results influenced the
interpretation of early numerical studies, and prompted many authors
to fit power-laws to the results of N--body simulations (Quinn et al
1986, Frenk et al 1988, Efstathiou et al 1988, Zurek, Quinn \& Salmon
1988, Crone et al 1994).  The general trends predicted by analytic
studies were generally confirmed, although significant deviations from
power-laws were also reported. These deviations were established
beyond doubt by the work of Dubinski \& Carlberg (1991) and Navarro,
Frenk \& White (1995), who found that halos formed in a Cold Dark
Matter (CDM) universe were best described by a density profile with a
gently changing logarithmic slope rather than a single power law.

\begin{figure}
\plottwo{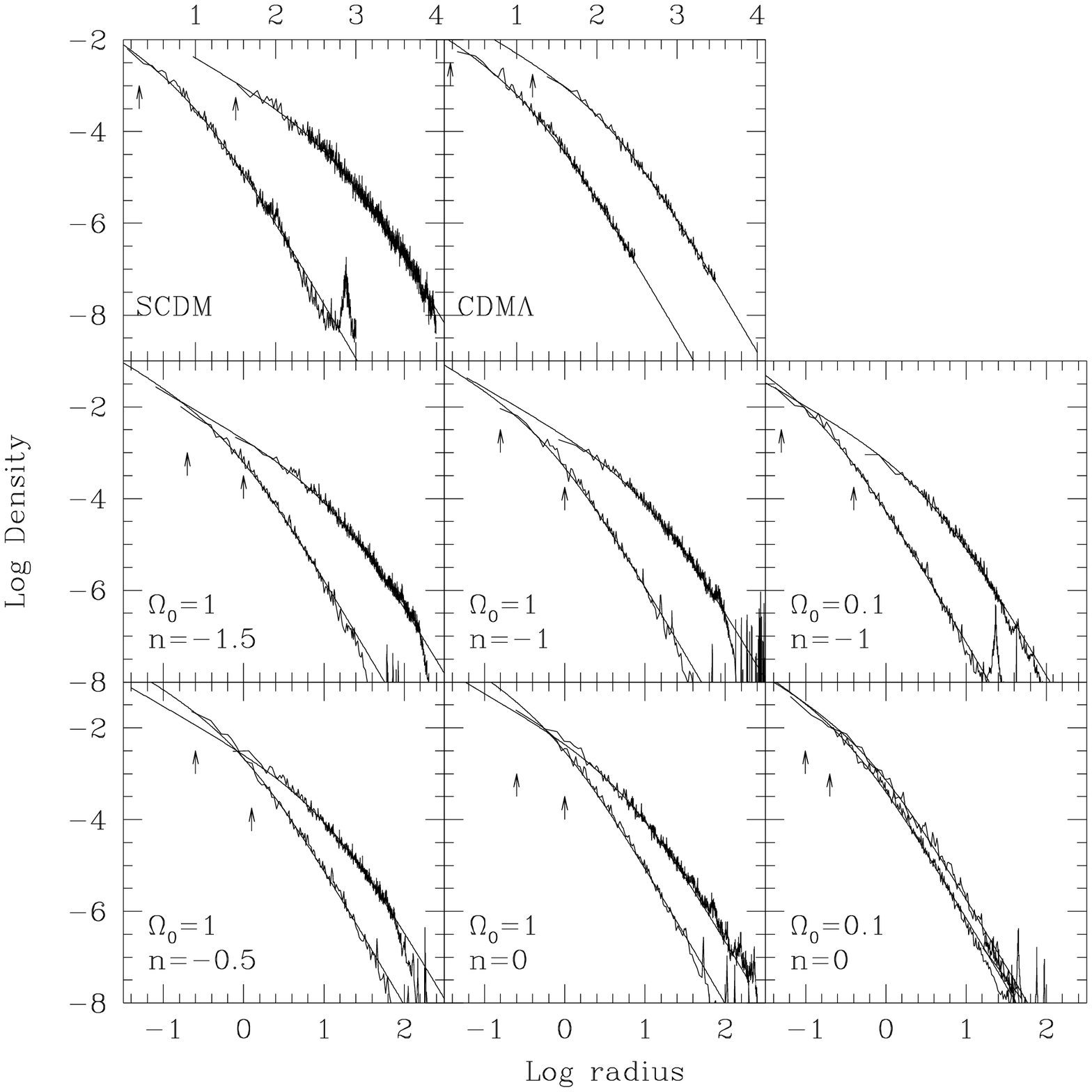}{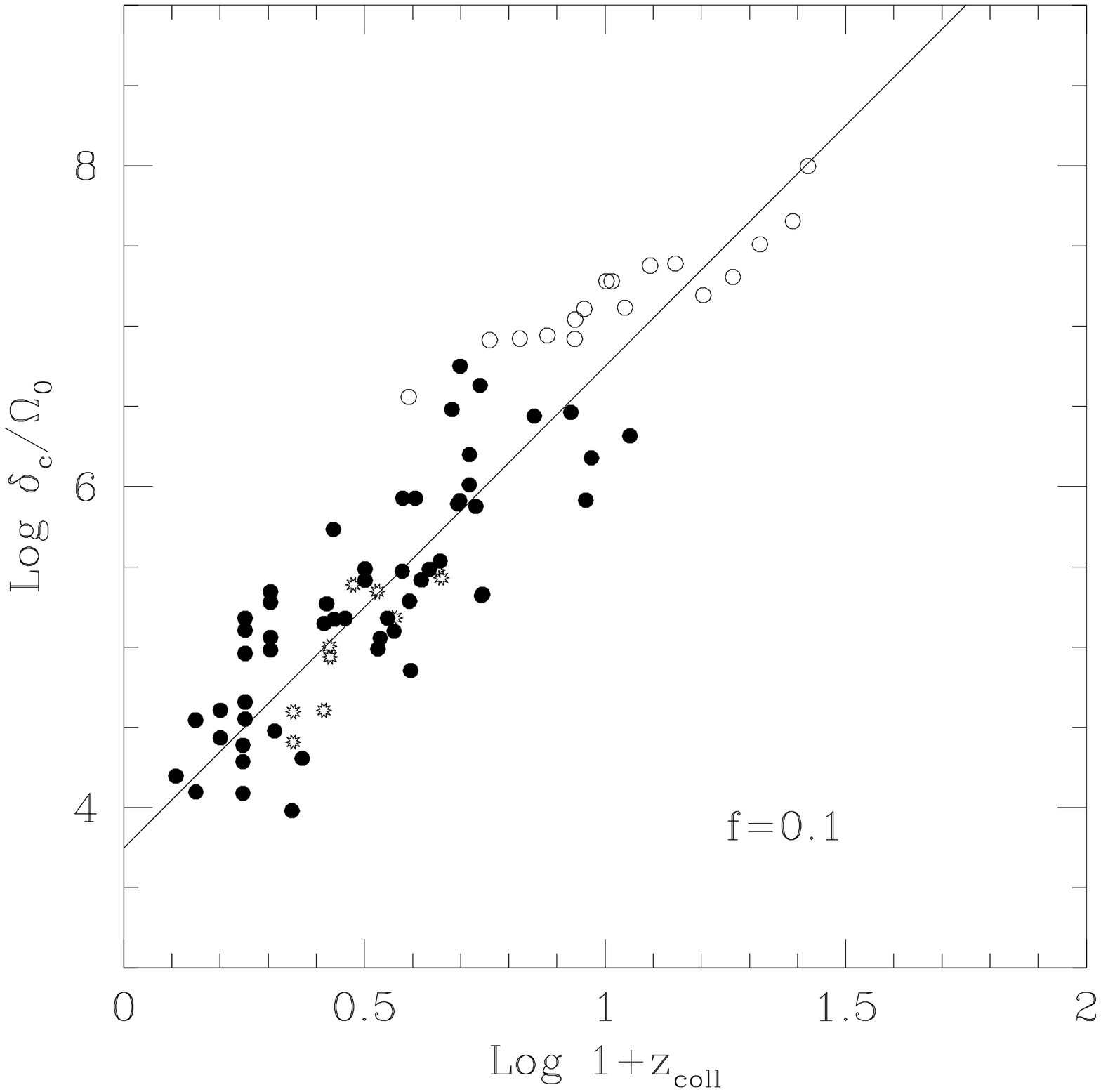}
\caption{
(a) Density profiles of dark matter halos. (b) Characteristic density
vs collapse redshift for all simulated halos measured directly in the
simulations. Filled circles refer to halos in $\Omega_0=1$ universes,
open circles correspond to those in open models and starred symbols to
the CDM$\Lambda$ model. The solid line shows the dependence predicted
by eq.~2.}
\label{fig-1}
\end{figure}

\section{A Universal Density Profile from Hierarchical Clustering}

Further simulations confirmed these results and indicated that this
structure appears universal: density profiles of halos of different
mass, formed in a variety of hierarchically clustering models (CDM and
power-law initial density fluctuation spectra, $P(k) \propto k^n$,
with different values of $\Omega_0$ and $\Lambda$), can be scaled to
look identical (Navarro, Frenk \& White 1997, NFW97).  This is shown
in Figure 1a, where we plot the spherically averaged density profiles
of one of the least and one of the most massive halos in each
series. These halos span four orders of magnitude in mass in the case
of the CDM models and about two orders of magnitude in mass in the
case of the power-law runs.

We define the mass of a halo, $M_{200}$, as that of a sphere with mean
interior density equal to $200 \rho_{crit}$, where
$\rho_{crit}=3H_0^2/8 \pi G$ is the critical density for closure. We
write Hubble's constant as $H_0=100\, h \, $km s$^{-1}$ Mpc$^{-1}$ in
this contribution. The radius of this sphere, $r_{200}$, is usually
called the ``virial radius'' of the halo. The virial radius and the
circular velocity at $r_{200}$, $V_{200}=(r_{200}/h^{-1} {\rm kpc})$
km/s, are alternative, equivalent measures of halo mass.

The solid lines in Figure 1a are fits of the form proposed by
Navarro, Frenk \& White (1996) [see also Cole \& Lacey (1996), and
Tormen, Bouchet \& White (1997)]
$$ {\rho(r) \over \rho_{crit}}= {\delta_c \over (r/r_s)(1+r/r_s)^2}.
\eqno(1)
$$
Here, $\delta_c$ is a (dimensionless) characteristic density, and
$r_s$ is a scale radius. 

This simple formula provides a good fit to the structure of
all halos over about two decades in radius, from the gravitational
softening radius (indicated by arrows in Figure 1a) to about the
virial radius of each halo. The quality of the fit is essentially
independent of halo mass or cosmological model, and implies a
remarkable similarity of structure between dark matter halos formed in
different hierarchically clustering scenarios.
\footnote{
We note that, although the density in eq.~1 diverges like $r^{-1}$
near the center, the simulations reported here do not prove that this
is the correct asymptotic behaviour. They only show that eq.~1
describes well the structure of halos in the radial range indicated
above. Recent work by Moore et al. and Kravtsov et al. (see their
contributions in this volume) suggests that the inner asymptotic slope
may differ from $r^{-1}$. Moore et al propose a steeper inner slope,
$r^{-1.4}$, for galaxy clusters and Kravtsov et al a shallower slope,
$r^{-0.7}$, for dwarf galaxies formed in CDM universes. Further
numerical work is underway to establish conclusively whether the inner
slope depends on mass in the way suggested by these authors.}

There is a single free parameter in eq.~1 for halos of given
mass. This parameter can be expressed either as the characteristic
density $\delta_c$ or as the ``concentration'' of the halo, defined by
the ratio $c=r_{200}/r_s$. ($\delta_c$ and $c$ are related by a simple
formula.) Our models show that $M_{200}$ and $\delta_c$ (or $c$) are
strongly correlated.  The characteristic density is simply
proportional to the mean density of the universe at the time of
collapse,
$$
\delta_c(M) \propto \Omega_0 (1+z_{coll}(M))^3. \eqno(2)
$$
as shown in Figure 1b. Hereafter we shall use the concentration, $c$,
as a measure of the characteristic density of a halo, since it is more
easily compared with observations.

Because collapse redshifts depend on the value of the cosmological
parameters, and can be computed analytically (see details in NFW97),
observational constraints on halo concentrations can be translated
directly into cosmological constraints. For example, the shape of the
power spectrum regulates the dependence of $c$ on halo mass. In
CDM-like models, where structure grows very fast, different mass
scales collapse almost at the same time, and $c$ depends weakly on
mass. Galaxy and galaxy cluster halos are thus expected to have values
of $c$ differing by a factor of about two, a useful prediction which
can be tested observationally. On the other hand, the mass dependence
of $c$ in a universe where structure develops slowly, e.g. a model
with white-noise initial perturbation spectrum ($P(k)=$constant),
would be much stronger.  Beyond the shape of the power spectrum, halo
concentrations can also be used to gain insight on the density of the
universe since, at fixed collapse redshift, the characteristic density
of a halo scales directly with $\Omega_0$ (eq.~2).  We investigate
below how rotation curves can be used to constrain the concentration
of dark halos surrounding disk galaxies and their consequences for CDM
models.

\section{Rotation Curves of Disk Galaxies}

\begin{figure}
\plottwo{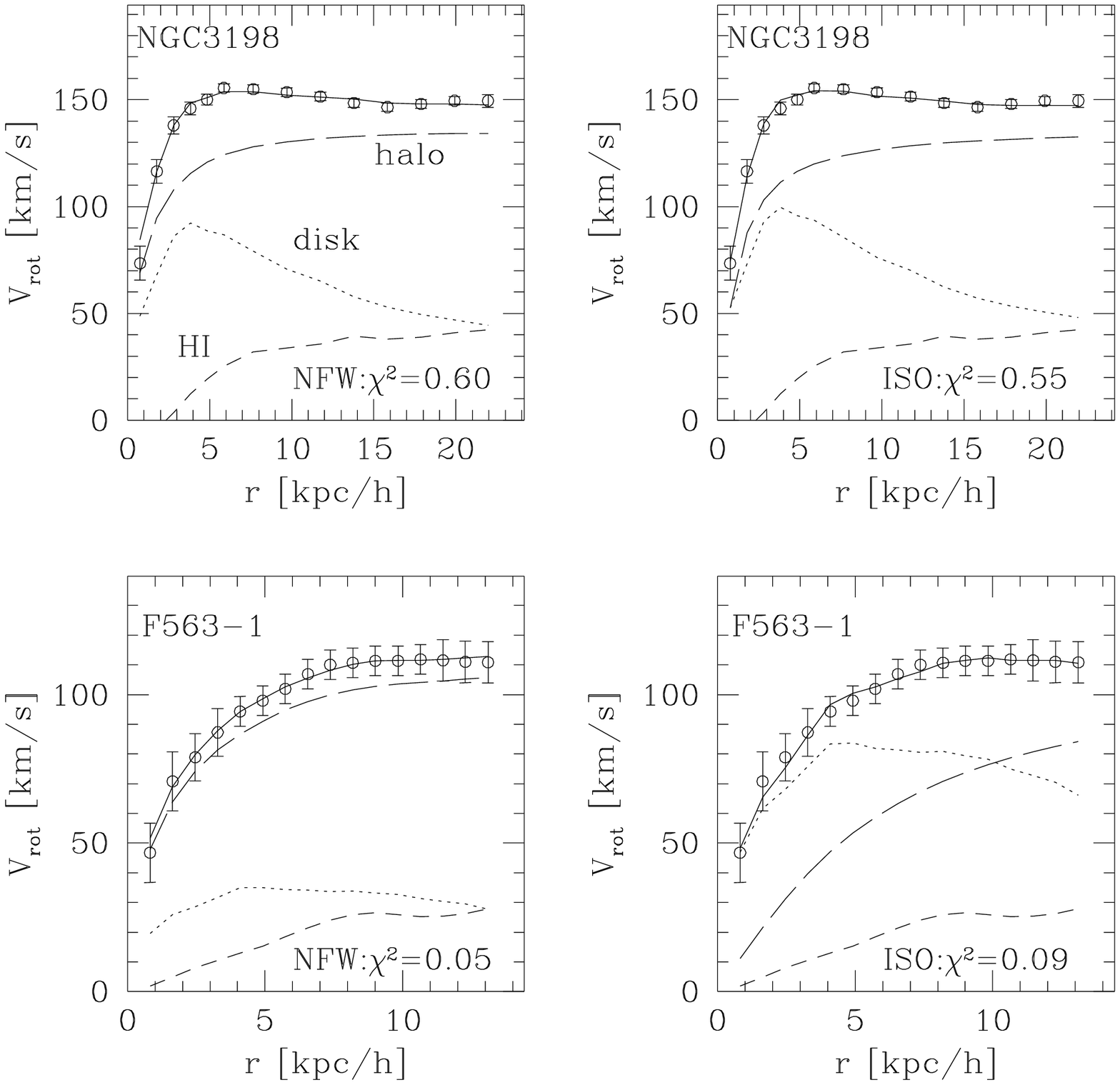}{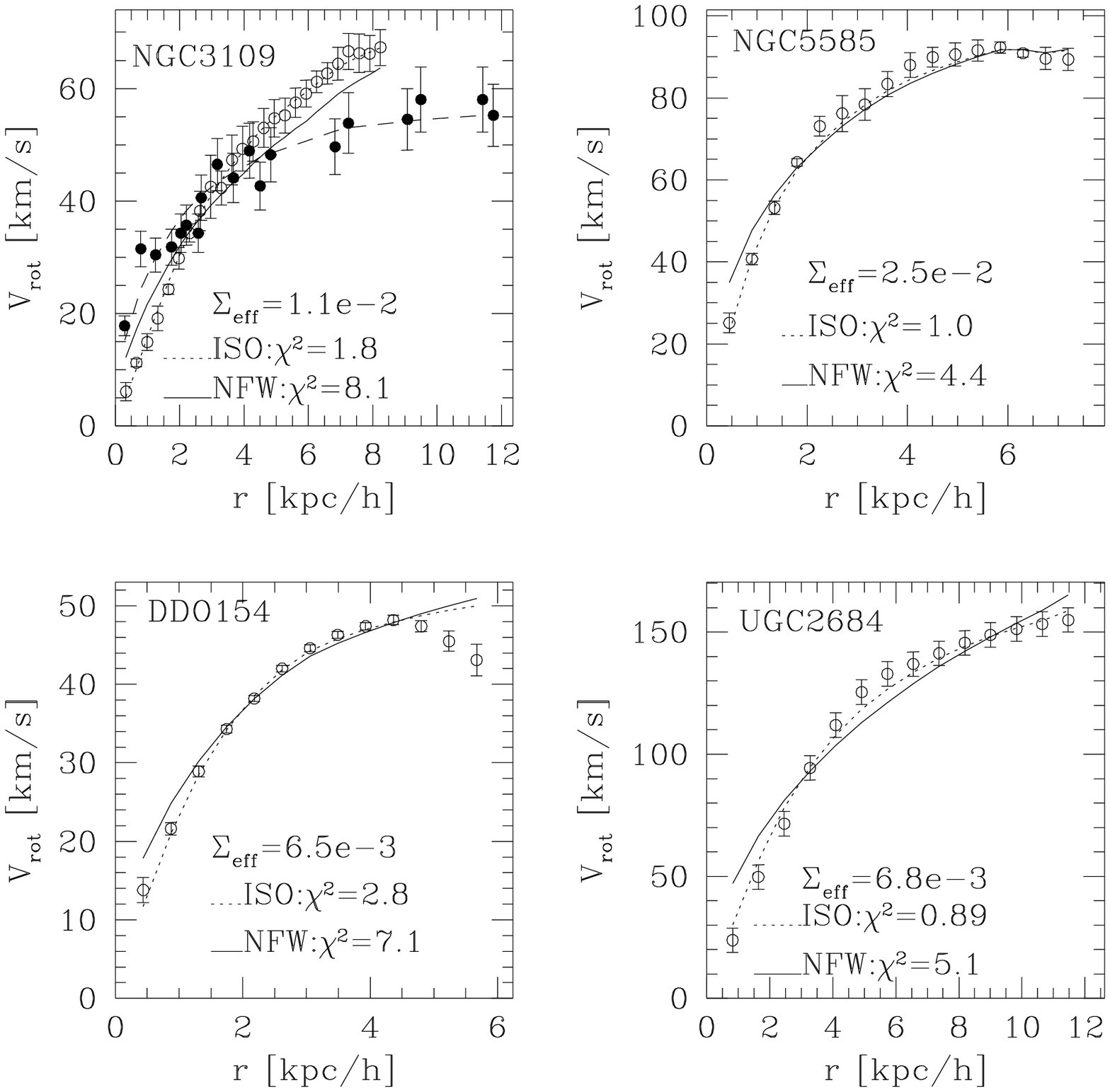}
\caption{
a)Rotation curve fits using the NFW and the ISO halo models shown for
a high-surface brightness galaxy (NGC 3198, Begeman 1987) and a
low-surface brightness galaxy (F563-1, de Blok 1997). Note that both
halo models produce acceptable fits, although they require different
disk mass-to-light ratios. b) same as (a), but for four LSB galaxies where
the ISO halo model fits better than NFW.}
\label{fig-2}
\end{figure}
%


We have analyzed the rotation curves of more than 100 disk galaxies
taken from the literature in order to examine whether the structure of
their surrounding dark halos is consistent with eq.~1 (for details,
see Navarro 1998). The sample covers a wide range in galaxy
parameters, spanning almost four orders of magnitude in luminosity,
two orders of magnitude in surface brightness, and almost a decade in
disk rotation speed. Figure 2a shows fits to the rotation curves of
two galaxies in the sample, performed using eq.~1 (hereafter called NFW
model) and a non-singular isothermal sphere (hereafter called ISO
model) for modeling the dark component. 

This figure serves to illustrate that rotation curves are generally
consistent with either NFW or ISO halo structures, although the
contribution of the disk to the circular velocity (dotted lines) can
differ dramatically depending on which model is adopted.  There are a
handful of exceptions: the HI rotation curves of six low-surface
brightness galaxies (LSBs) are better fitted with an ISO model. Four
of these galaxies are shown in Figure 2b. The rotation speed seems
indeed to rise too rapidly with radius (as a ``solid body'') to be
consistent with the NFW mass profile. However, the differences are
small, and the significance of the discrepancy may have been
overemphasized by optimistic velocity error bars.

The case of NGC 3109 shows that this is a true possibility. Here, two
datasets are available for the same galaxy, one based on HI
observations only (open circles, Jobin \& Carignan 1990), and one
based on independent H$\alpha$ and HI observations (filled circles,
Carignan 1985). The two resulting rotation curves are dramatically
different. While the open-circle data excludes the NFW model with high
significance, the filled circles are fully compatible with an NFW halo
profile.  Therefore, the possibility remains that the discrepancies
between NFW halo models and the rotation curves of some low-surface
brightness galaxies may be less important than previously thought
(Moore 1994, Flores \& Primack 1994). A thorough reanalysis of the
rotation curves of these galaxies is needed to sort out the problem.
Other possible resolutions of this problem include modifications to
the dark matter profile caused during the assembly of the disk (Navarro,
Eke \& Frenk 1996), or small systematic deviations from a strict NFW
shape, as proposed by Kravtsov et al (1997).


%
\begin{figure}
\plottwo{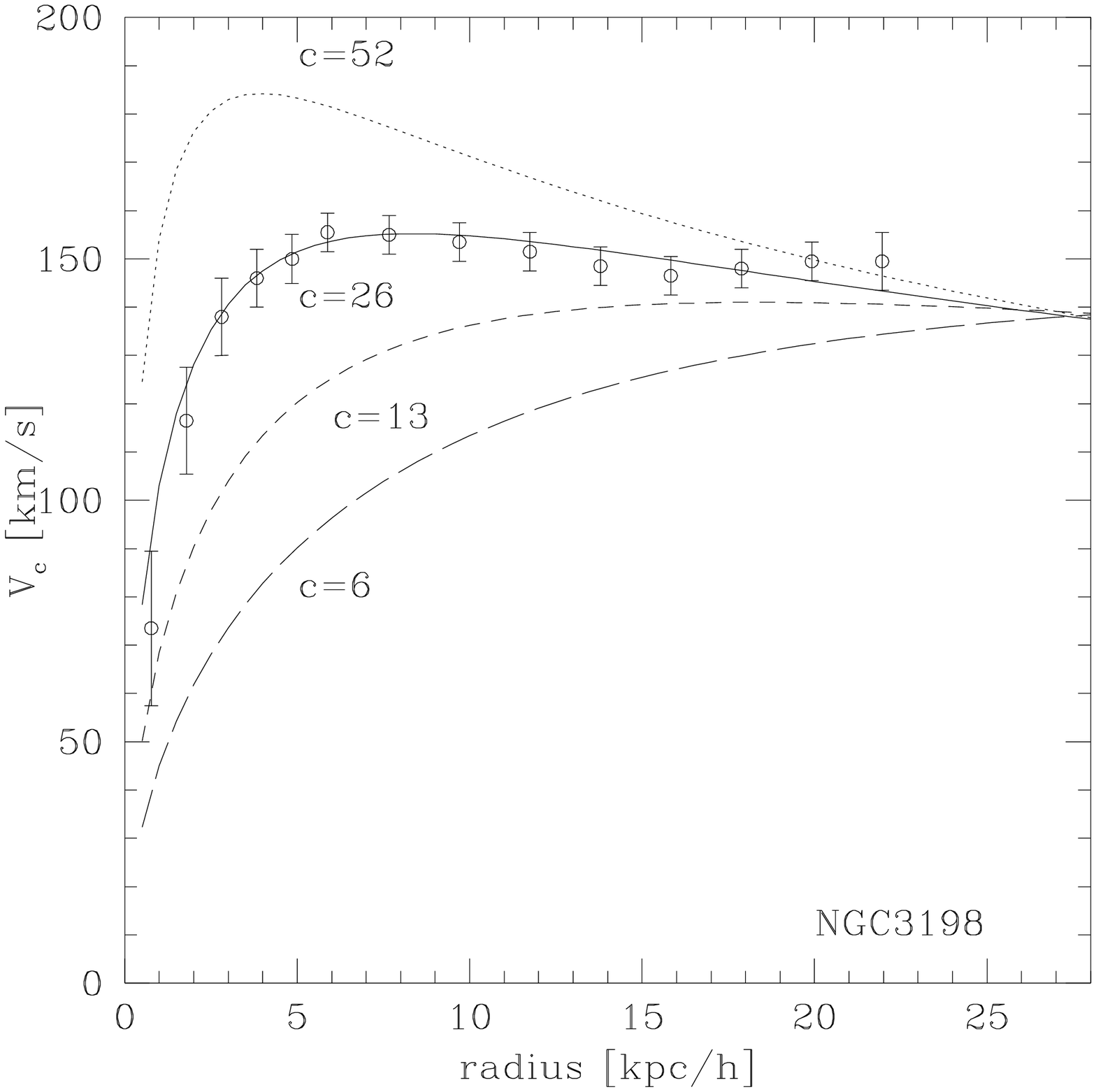}{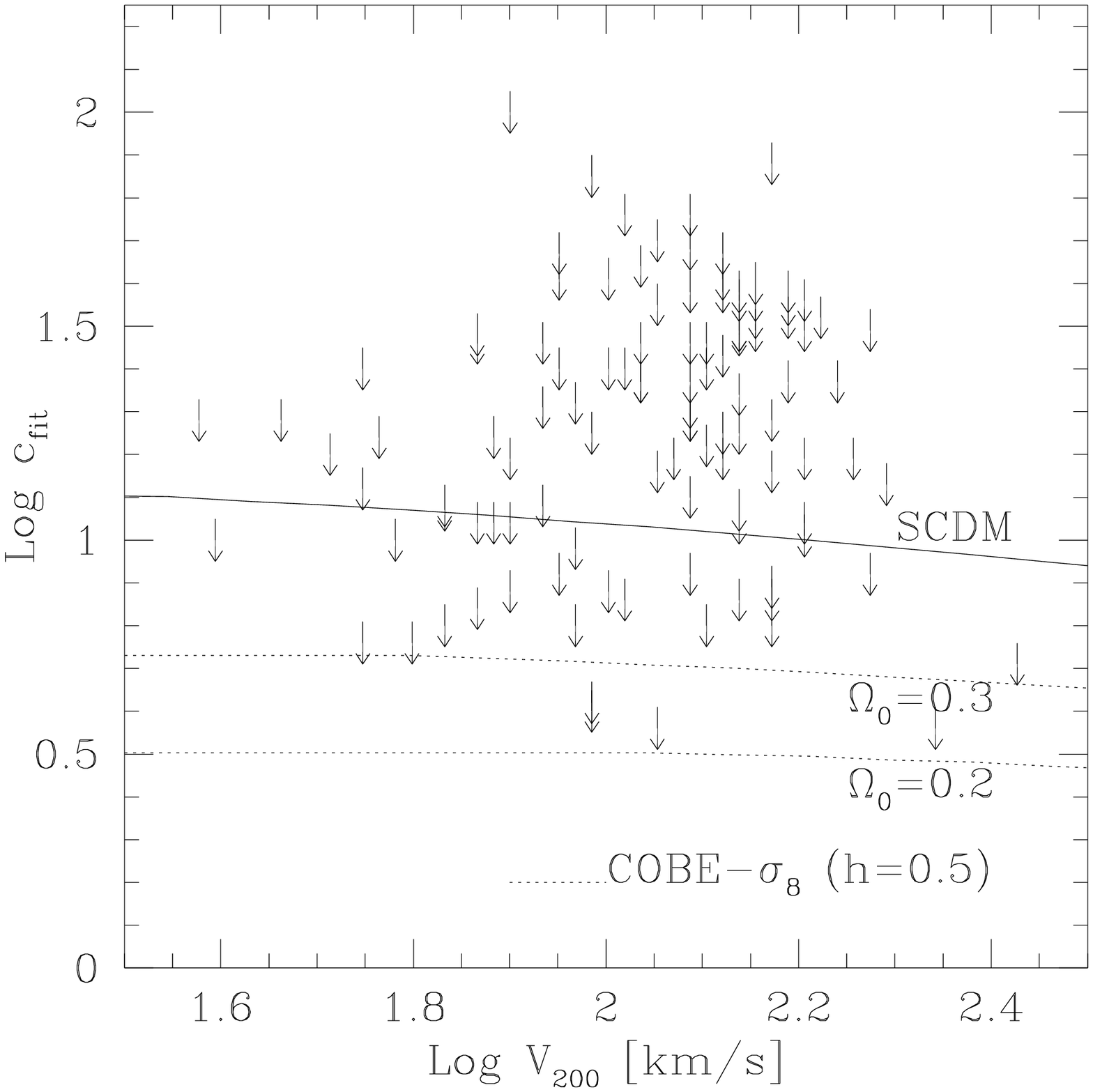}
\caption{
(a){\it Halo-only} circular velocity curves shown for different values
of the concentration, $c$, and similar values of $V_{200}$, and
compared with rotation curve data for NGC 3198. (b) Upper limits on
halo concentration, $c_{fit}$, computed individually for all galaxies
in the sample, shown as a function of the fitted $V_{200}$.}
\label{fig-3}
\end{figure}

With this caveat, let us now explore the consequences for CDM models
of fitting NFW halos to disk galaxy rotation curves.The first thing to
note is that the overall shape of the rotation curve defines a firm
{\it upper bound} to the concentration of the halo. This is shown in
Figure 3a, where we illustrate the circular velocity profiles of halos
of similar mass (ie. similar $V_{200}$) but different
concentrations. These curves ignore the contribution of the luminous
component.  From the figure, it is clear that $c=26$, the value of the
concentration that best fits the rotation curve neglecting the
luminous component, represents an upper limit to the concentration of
the halo that surrounds NGC 3198. Halos with $c<26$ could in principle
be made consistent with the data by suitable addition of a massive
disk component, but $c>26$ halos result in rotation speeds that are in
excess of the data even before allowing for the presence of the
disk. This upper limit is quite insensitive to the halo mass adopted
(expressed by $V_{200}$), which merely sets the velocity scale of the
fit.

A second thing to note in Figure 3a is that the value of $c$ retrieved
by fitting {\it halo-only} models to the data (referred to hereafter
as $c_{fit}$) is a good indicator of the {\it shape} of the rotation
curve.  Values of $c_{fit} \, \lsim \, 10$-$20$ indicate that the
rotation curve rises slowly, while $c_{fit} \, \gsim \, 10$-$20$
describe a sharply rising rotation curve that is either flat or
declines in the outer regions.

Upper limits to the halo concentration, ie. $c_{fit}$, derived
individually for all galaxies in our sample are shown in Figure 3b as
a function of $V_{200}$. Overlaid are halo concentrations expected in
three CDM cosmogonies.  SCDM refers to the standard biased
($\sigma_8=0.6$) $\Omega=1$ CDM model. The two dotted lines correspond
to low-density, flat ($\Omega+\Lambda=1$) CDM cosmogonies normalized
to match the fluctuations in the cosmic microwave background observed
by COBE (see, eg., Eke, Cole \& Frenk 1996). The scatter around each
of these lines is expected to be less than about $30 \%$ (Navarro et
al, in preparation). As discussed earlier, the lines are almost
horizontal, indicating that the concentration is a weak function of
mass in CDM models.  The data presented in Figure 3b seems to be
inconsistent with the SCDM model, and appears to favor the low-density
CDM models since, in order to be acceptable, halo concentrations
should be below all individual upper limits.

It is instructive to see which galaxies cannot be reconciled with the
SCDM model. Figure 4 shows $c_{fit}$ as a function of the effective
surface brightness of the galaxy, $\Sigma_{eff}=L_I/\pi r_{disk}^2$,
defined as the surface brightness of a galaxy if all its light were
concentrated within one exponential disk scalelength. (All
luminosities quoted are in the I-band.) There is a strong correlation
between surface brightness and $c_{fit}$, indicating that LSBs have
slowly rising rotation curves while high-surface brightness galaxies
(HSBs) have steeply rising, flat rotation curves.  Most galaxies
incompatible with SCDM (ie, those with $c_{fit} \, \lsim \, 10$) are
LSBs.

What does the correlation between $c_{fit}$ and $\Sigma_{eff}$ mean?
If the universe is dominated by cold dark matter, we expect all halos,
regardless of mass, to have similar values of $c$ previous to the
collapse of the luminous component (see the nearly horizontal lines in
Figure 3b). Thus, the $c_{fit}$-$\Sigma_{eff}$ correlation reflects
the relative importance of disks in shaping the rotation curves of
galaxies of different surface brightness. In very low surface
brightness systems the luminous component is unimportant
gravitationally and the rotation curve traces the mass distribution of
the halo, ie. $c_{fit} \sim c \approx 3$ or $5$, depending on the
value of $\Omega_0$ (see Figure 3b). In high surface brightness
systems the disk gravity steepens the rotation curve in the inner
regions, resulting in higher values of $c_{fit}$.

\begin{figure}
\plotfiddle {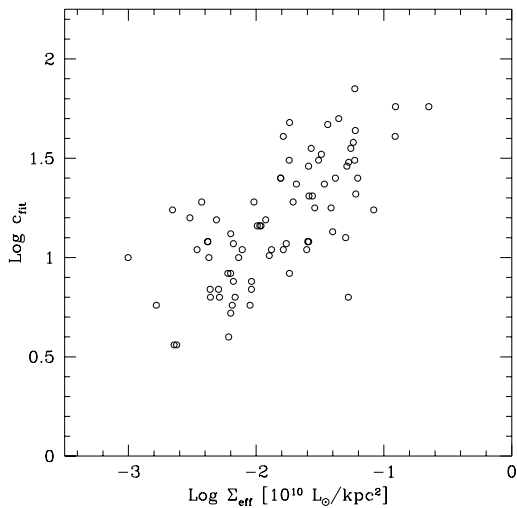} {150pt} {0} {35}{35} {-100}{-55}
\caption{The shape of the rotation curve, parameterized by $c_{fit}$, plotted as a function of effective surface brightness. }
\label{fig-4}
\end{figure}
%


We can take this analysis one step further and ask what the
$c_{fit}$-$\Sigma_{eff}$ correlation means for the mass-to-light ratio
of the disk and for the relationship between the halo circular
velocity and the rotation speed of the disk. In other words, assuming
that all halos have initially the same concentration, eg. $c \approx
3$ (for $\Omega=0.2$, $\Lambda=0.8$, see Figure 3b), for what
combination of halo masses and disk mass-to-light ratios can one
recover the observed $c_{fit}$-$\Sigma_{eff}$ relation whilst at the
same time satisfying other constraints such as the luminosity-surface
brightness relation and the Tully-Fisher relation?

For illustration, let us assume that all disks have the same stellar
mass-to-light ratio, $(M/L)_{disk}=1 h
(M_{\odot}/L_{\odot})$. Assuming that the initial concentration of the
halo is $c=3$, we compute $c_{fit}$ for disks of different surface
brightness. At each surface brightness we take the luminosity of the
disk to be that given by the $L_I$-$\Sigma_{eff}$ relation. We then
adjust the circular velocity of the halo ($V_{200}$) in order to match
the rotation speed within the luminous radius of a galaxy of that
luminosity given by the Tully-Fisher relation. Both observational
relations are constructed internally using galaxies in the same
sample.

The result of this exercise is shown with short-dashed lines in Figure
5. Halos are required to be more massive than expected from the disk
rotation speed (ie. $V_{200}>V_{rot}$) in order to satisfy the
Tully-Fisher relation (upper-right panel). However, no significant
correlation is found between surface brightness and the shape of the
rotation curve; irrespective of surface brightness all galaxies have
approximately the same value of $c_{fit}$ (upper-left panel).

A second example is provided by the dotted lines in Figure 5, which
assume that the circular velocity of the halo is the same as the
rotation speed of the disk, ie. $V_{200}=V_{rot}$. In this case the
mass-to-light ratio of the disk has to be higher than unity to
match the Tully-Fisher relation. As is clear from Figure 5, this
assumption also provides a poor fit to the $c_{fit}$-$\Sigma_{eff}$
relation.

Thus, the existence of a correlation between $c_{fit}$ and
$\Sigma_{eff}$ implies that the disk mass-to-light ratios and the
ratio between $V_{200}$ and $V_{rot}$ cannot remain constant for all
galaxies. The solid and long-dashed lines in Figure 5 are constructed
to match the observed $c_{fit}$-$\Sigma_{eff}$ relation. (Solid and
long-dashed lines refer to halos formed in the $\Omega=0.2$ and
$\Omega=0.3$ models shown in Figure 3b, respectively.) The resulting
disk mass-to-light ratios increase from $\sim 0.5$ in faint,
slow-rotating disks to $3$-$5 \, h M_{\odot}/L_{\odot}$ in the fastest
rotators: $(M/L_I)_{disk} \approx (L_I/10^9 L_{\odot})^{0.3} h
\, M_{\odot}/L_{\odot} \approx (V_{rot}/100 \, $\kms$) \,
M_{\odot}/L_{\odot} $. The color differences between disks of
different morphology/surface brightness suggest that systematic trends
of this magnitude between disk mass-to-light ratios and luminosity do
indeed exist (de Jong 1995).

The relationship between $V_{200}$ and $V_{rot}$ that results is also
intriguing. Halos of disks with $V_{rot} < 150 \, $ \kms have $V_{200}
> V_{rot}$, by up to $60 \%$ for $V_{rot} \sim 100$ \kms. On the other
hand, disks that rotate faster than $\sim 150$ \kms all have similar
halo circular velocities, $V_{200} \approx 200$ \kms. This is
reminiscent of a well-known result of dynamical studies of
satellite/primary pairs: there is little correlation between the
rotation speed of luminous disks and the mass of their surrounding
halos (Zaritsky et al 1997). It is comforting that we arrive at a
similar conclusion using a completely different approach.

One curious corollary is that few disk galaxies inhabit halos more
massive than $V_{200} \approx 200$ \kms, a result which may reflect
the onset of disk instabilities in massive galaxies. As discussed by
Mo, Mao \& White (1997), stable disks embedded in NFW halos are only
stable if the disk contributes a small fraction of the total
mass. More specifically, their analysis suggests that only in systems
where $M_{disk}/M_{200} < \lambda$ can disks avoid being disrupted by
global instabilities. Here $\lambda=J E^{1/2}/GM^{5/2}$ is the usual
dimensionless rotation parameter which, as shown by extensive
numerical work, seldom exceeds $\sim 0.1$ (Cole \& Lacey 1996).  In
other words, if the amount of baryons that collapse to form the disk
is such that $M_{disk}/M_{200}$ exceeds about $0.1$, very few of these
systems would survive as disks to the present.  The fraction of the
total mass that can collect in the disk cannot exceed the universal
baryon fraction, $\Omega_b/\Omega_0$ (White et al 1993). For the low
density models we are considering here, and adopting the usual Big
Bang nucleosynthesis value for $\Omega_b$, $M_{disk}/M_{200} \, \lsim
\, 0.2$-$0.3$. We see that long-lived disks can only form in systems where
fewer than about half of all available baryons have cooled and
assembled into the disk.

The lower-right panel in Figure 5 shows that, in models that satisfy
the observed $c_{fit}$-$\Sigma_{eff}$ relation, the disk mass fraction
increases sharply with halo mass, exceeding the critical value of
$\sim 0.1$ at $V_{200} \approx 200$ \kms. Matching the rotation curve
shapes thus requires the mechanism regulating the disk mass fraction
(e.g. feedback from supernovae and evolving stars) to be highly
efficient in low mass halos, but relatively inefficient in halos more
massive than about $V_{200} \approx 200$ \kms.  Indeed, this rapidly
varying ``efficiency'' of assembly of baryons into galaxies is at the
heart of all successful hierarchical galaxy formation models, where it
is invoked to reconcile the relative scarcity of dwarf galaxies with
the myriad of low-mass halos expected in hierarchically clustering
universes (Kauffmann, White \& Guiderdoni 1993, Cole et al 1994). It
is interesting and suggestive that the same feedback process needed to
explain the relative number of dwarf and bright galaxies in
hierarchical models is actually required to match the shape of the
rotation curves of present-day disk galaxies.

\begin{figure}
\plotfiddle {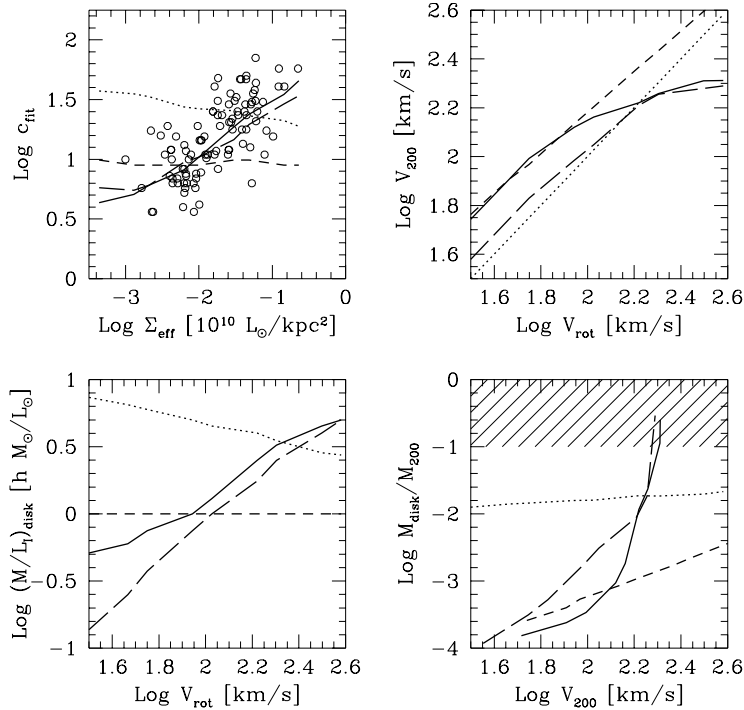} {225pt} {0} {50}{50} {-150}{-90}
\caption{
Correlations between different parameters of various different
rotation curve models, described in detail in the text.}
\label{fig-5}
\end{figure}
%


\acknowledgments

Much of the work described here is part of a collaboration with Carlos
Frenk and Simon White. I am grateful to Tim Pickering, Paolo Salucci,
Erwin de Blok, and Liese van Zee for providing data in electronic
form.

\end{document}